\documentclass[aps,pre,reprint,groupedaddress,showpacs]{revtex4-1}

\usepackage[english]{babel}
\usepackage[latin2]{inputenc}
\usepackage{graphicx}
\usepackage{amsmath}

\begin{document}
\title{Non-Gaussian polymers described by alpha-stable chain statistics: model, applications and effective interactions in binary mixtures.}

\author{M. Majka}
\email{maciej.majka@uj.edu.pl}
\affiliation{Marian Smoluchowski Institute of Physics, Jagiellonian University, ul. prof. Stanis\l{}awa \L{}ojasiewicza 11, 30-348 Krak\'{o}w Poland}
\author{P. F. G\'{o}ra}
\affiliation{Marian Smoluchowski Institute of Physics, Jagiellonian University, ul. prof. Stanis\l{}awa \L{}ojasiewicza 11, 30-348 Krak\'{o}w Poland}

\begin{abstract}
The Gaussian chain model is the classical description of a polymeric chain, which provides the analytical results regarding end-to-end distance, the distribution of segments around the mass center of a chain, coarse grained interactions between two chains and effective interactions in binary mixtures. This hierarchy of results can be calculated thanks to the alpha stability of the Gaussian distribution. In this paper we show that it is possible to generalize the model of Gaussian chain to the entire class of alpha stable distributions, obtaining the analogous hierarchy of results expressed by the analytical closed-form formulas in the Fourier space. This allows us to establish the alpha-stable chain model. We begin with reviewing the applications of Levy flights in the context of polymer sciences, which include:  chains with heavy-tailed distributions of persistence length, polymers adsorbed to the surface and the chains driven by a noise with power-law spatial correlations. Further, we derive the distribution of segments around the mass center of the alpha-stable chain and the coarse-grained interaction potential between two chains is constructed. These results are employed to discuss the model of binary mixture consisting of the alpha-stable chains. On what follows, we establish the spinodal decomposition condition generalized to the particles described by the shape of alpha-stable distributions. This condition is finally applied to analyze the on-surface phase separation of adsorbed polymers, which are known to be described with heavy tailed statistics.
\end{abstract}


\maketitle

\section{Introduction}
While the theory of Flory provides an accurate description of the ideal polymeric chains \cite{bib:teraoka}, factors such as complex environment interactions, adsorption or designed chemical composition can lead to significant deviations from this model. The Flory approach is based on the Gaussian chain model, in which the conformation of a chain is equivalent to the trajectory of a particle undergoing the thermal Brownian motion \cite{bib:teraoka}. In this model the chain is characterized by the Gaussian distribution of the nearest neighbor distances, a fact that leads to the entire hierarchy of analytical results. In particular, the Gaussian shape propagates to such characteristics as: end-to-end distance distribution \cite{bib:teraoka}, distribution of segments around the mass center of the chain \cite{bib:debye} and the coarse-grained interaction potential between two chains in terms of the distance between their mass centers \cite{bib:flory}. Deriving all of these characteristics is possible due to a single fact: the Gaussian distribution is alpha-stable. Since there exist the entire class of alpha-stable, heavy-tailed distributions \cite{bib:samorodnitsky}, this suggests that a natural and equally prolific generalization of the Gaussian chain model can be based on the alpha-stable distributions.  Indeed, in this paper we discuss the alpha-stable chain model and calculate all of the characteristics analogous to the Gaussian model.

The first goal of this paper is to establish the physical context in which the alpha stable distributions are relevant for the polymer sciences. Since the application of alpha-stable distributions (or Levy walks/flights) in this context is not an entirely new concept, in Section \ref{sec:interpretation} we review the relevant literature. In addition, we provide our own simulations of a model polymeric chain under the spatially correlated  noise, which establish another context for our considerations.

The main part of this paper is focused on deriving and analyzing the different aspects of the alpha-stable chain model. In Section \ref{sec:model} we introduce the model itself. In Section \ref{sec:mass_center} the distribution of nodes around the mass center of a chain is calculated and in Section \ref{sec:interaction} the coarse-grained model of interaction between two chains is established. All of these results are analytical and closed-form in the Fourier space.  

Another goal of this paper is to analyze the stability of binary mixtures composed of the alpha-stable chains. Understanding the behavior of binary mixtures is a vital problem in industry, medicine, wet nano-technology and biophysics. The stability of solution is governed by effective interactions \cite{bib:lekkerkerker}, predicting which is a classical problem of soft matter physics \cite{bib:likos}. In the context of binary mixtures of Gaussian particles, Bolhuis et al. found via simulations that the interaction between particles of one species has also a Gaussian profile, but with an addition of a shallow attractive tail \cite{bib:bolhuis}. Similar results were predicted half-analytically via closure-relations techniques in \cite{bib:gauss1,bib:gauss2}. On the other hand, a simpler, but entirely analytical method has been recently proposed by the present authors, in \cite{bib:eff_int}. Therein, we have studied the stability of Gaussian particles mixtures and our results proved similar to the spinodal decomposition analysis \cite{bib:sep1,bib:sep2}, namely, there exist a well defined region of mixing and demixing, dependent on the proportion of gyration radii describing the two types of particles. Since our methodology from \cite{bib:eff_int} can be conveniently extended to the particles described with the alpha-stable profiles, we apply this approach in the current paper. As the main result of Section \ref{sec:bin_mix}, we generalize the spinodal decomposition condition for Gaussian particles to the entire class of particles based on the alpha-stable distributions. We discuss the validity of our methodology in Section \ref{sec:discussion}.

Finally, in Section \ref{sec:gauss_sep} we employ the results regarding phase separation in binary mixtures to analyze the phase separation of adsorbed polymers versus their behavior in the bulk. As a result we predict the parameters for which the different combinations of simultaneous mixing/demixing on the surface and in the bulk can be achieved.

\section{Levy flights in polymer sciences}\label{sec:interpretation}
Except for the Gaussian case, the asymptotic behavior of the alpha stable distributions is of the power-law type \cite{bib:samorodnitsky}. A random walk characterized by such a heavy-tailed distribution of steps is known as Levy flight. It is usually difficult to interpret Levy flights in physical terms, thus let us discuss three situations justifying such statistics in the context of polymers.

The first scenario can be related to the non-Gaussian distribution of segment persistence lengths. The Gaussian chain model is usually interpreted as a continuous limit of a discrete model, in which all segments have the same persistence length \cite{bib:teraoka}. However, it can be also seen as the model for chain made of unequal segments, whose persistence lengths follow the Gaussian distribution. This can be further generalized to the alpha-stable distribution, the idea suggested by Moon and Nakanishi in \cite{bib:moon}. They proposed the Levy walk chain model, based on the formalism of turbulent flows \cite{bib:turbulence} and calculated Flory exponents for this model. While no direct experimental confirmation of this idea is known to authors, non-Gaussian persistence length distribution might be the result of the varying chemical composition of a chain. E. g. DNA double strand is characterized by the persistence length much greater than a single base pair \cite{bib:DNAstrand}, but also certain sequences of chemical monomers can assemble into relatively long and stable structures of significant persistence length, such as protein domains \cite{bib:AFMdomains}. A possible realization of such 'Levy flights' could be the intrinsically disordered proteins, in which second-order structural motifs such as alpha-helices coexist with disordered loops \cite{bib:disordered1, bib:disordered2}. However, it should be mentioned that some numerical experiments on the structure of partially unfolded proteins indicate that Gaussian statistics is rather robust \cite{bib:fitzkee}. 

Another scenario is similar to the problem of a tracer which mixes one- and three-dimensional diffusion. Such motion has been observed experimentally in DNA-binding proteins \cite{bib:DNA1} and its simulations revealed the heavy-tailed distributions of steps along the polymer in certain configurations \cite{bib:DNA2}. This behavior can be efficiently modeled with Levy flights \cite{bib:DNA3}. In the context of polymers, we consider the adsorption of a chain to the surface. This problem was first analyzed by de Gennes from the scaling perspective in \cite{bib:degennes1, bib:degennes2}. For the intermediate attraction strength, only some fraction of segments is attached to the surface, while the loops that connect those segments diffuse into the bulk. Considering the projection of the chain on the surface, it has been argued by Bouchaud and Daoud \cite{bib:bouchaud} that the planar trajectory connecting adsorbed nodes can be modeled as Levy flights, since the subsequent adsorbed segments connected by a loop can be found at abnormally long distances. Bouchaud and Daoud predict that for an adsorbed polymer its gyration radius parallel to the surface scales as $R_{g,||}\propto N^{3/4}$ for the strong adsorption and $R_{g,||}\propto N^{3/5}$ for the weak adsorption \cite{bib:bouchaud}. Within the Flory-type theory for Levy flights discussed in \cite{bib:bouchaud}, the former translates into the characteristic exponent of the distribution $\alpha=1$ and the latter demands $\alpha=2$. This means that while in the strong adsorption regime the Levy flights and the Gaussian statistics are equivalent, for the weak adsorption limit $R_{g,||}$ should be modeled with power-law distributions. 

One final interpretation can be related to the situation in which a polymer experiences a random, though spatially correlated behavior of surrounding environment. Such conditions occur in the glassy state, in which the correlations are exponential \cite{bib:glass1, bib:glass2, bib:glass3}, or near crystallization, in which case the scale-free behavior results in the power law correlations \cite{bib:binney}. In \cite{bib:majka1} we have simulated a two dimensional model polymeric chain driven by the spatially correlated noise and observed the effect of spontaneous chain unfolding, i.e. a significant number of segments tends to form linearized structures, scattered along the chain. As we have shown in \cite{bib:majka2}, this effect was mainly due to the short (2-3 segments) structures, but structures up to 50 segments were also observed. Such elongated fragments may act as the Levy flights, provided that their distribution is wide enough. 

\begin{figure*}
\includegraphics[width=0.98\textwidth]{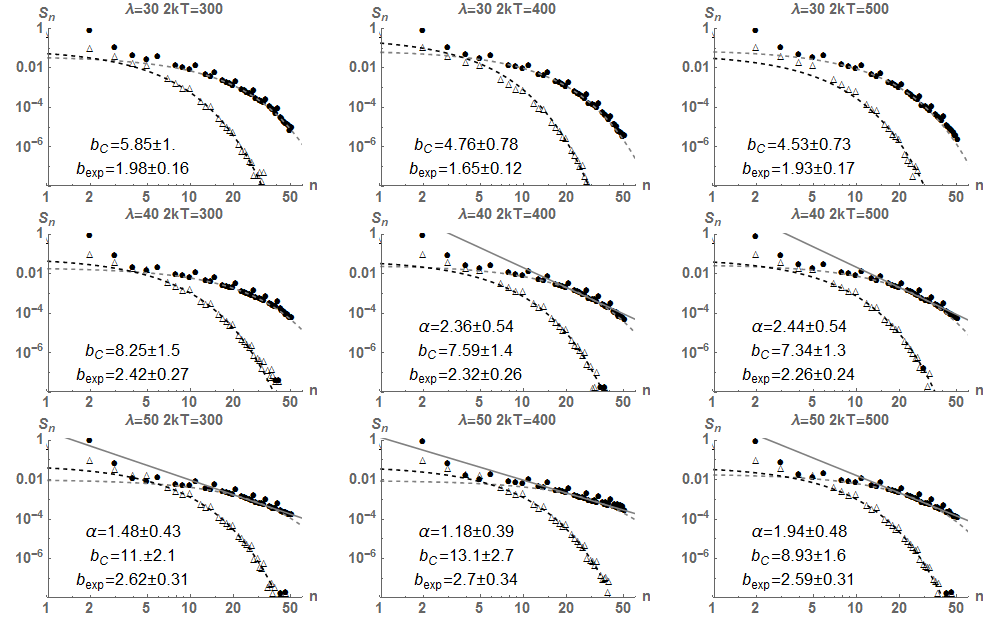}
\caption{The distribution of linearized polymer structures $S_n$ in the system driven by the spatially correlated noise. Each plot represents the results for noise amplitude $2kT$ and correlation length $\lambda$. Triangles: exponential noise correlation function $<\boldsymbol{\xi}(\textbf{r})\boldsymbol{\xi}(\textbf{r}+\Delta\textbf{r})>\propto\exp(-|\Delta\textbf{r}|/\lambda)$, dots: Cauchy noise correlation function $<\boldsymbol{\xi}(\textbf{r})\boldsymbol{\xi}(\textbf{r}+\Delta\textbf{r})>\propto1/\left(1+(|\Delta\textbf{r}|/\lambda)^2\right)$. The tail behavior ($n\ge17$) has been fitted with exponential decay model $S_n=ce^{-n/b_i}$ (dashed lines) with $b_i$ given on the plot. For $\lambda\ge40$ the data has been also fitted with power-law $S_n=cn^{-(\alpha+1)}$ for $n\ge17$, indicating the power law asymptotic behavior.}\label{fig:linearized}
\end{figure*}

The model from \cite{bib:majka1} consisted of a bead-spring chain with a global Lennard-Jones potential assigned to each bead and a second-nearest neighbor harmonic interaction to induce non-linear conformations. The system was driven by the noise $\boldsymbol{\xi}$, which spatial correlation function read:
\begin{equation}
<\boldsymbol{\xi}(\textbf{r})\boldsymbol{\xi}(\textbf{r}+\Delta\textbf{r})>\propto\exp(-|\Delta\textbf{r}|/\lambda)
\end{equation}
For the purpose of the current paper we have repeated the simulations from \cite{bib:majka1}, replacing the exponential correlations in the noise with a heavy-tailed function, based on the Cauchy distribution, namely:
\begin{equation}
<\boldsymbol{\xi}(\textbf{r})\boldsymbol{\xi}(\textbf{r}+\Delta\textbf{r})>\propto1/\left(1+(|\Delta\textbf{r}|/\lambda)^2\right)
\end{equation}
The data regarding the linearized fragments has been gather in the same fashion as in \cite{bib:majka2}. To improve statistics, the single set of parameters was simulated 128 times, otherwise, the details of the simulations remained the same as in \cite{bib:majka1} and \cite{bib:majka2}. In Fig. \ref{fig:linearized}, we include the representative probability distributions $S_n$ of finding the $n$-segments long structure in the chain geometry. The data has been gathered from the regime of noise dominated dynamics. With the growing correlation length $\lambda$ the distribution $S_n$ gradually develops a linear region in the log-log plot. For $\lambda\ge40$, where relevant, we have fitted $S_n$ for $n>17$ with a power-law model $S_n=cn^{-(\alpha+1)}$. We obtain $\alpha$ ranging from $1.18\pm0.39$ to $2.44\pm0.54$, with the relative error typically at the level of 20-30\%. The uncertainty intervals for these values of $\alpha$ overlap with the interval $\alpha\in[0,2]$, which is expected for the asymptotic behavior of the alpha-stable distributions \cite{bib:samorodnitsky}. For comparison, the data have been also fitted with the exponential decay model $S_n=ae^{-n/b}$. While both the linear and exponential fit describe the tail part of $S_n$ similarly well (in both cases $R^2\simeq 0.8$) and, most probably, even for $\lambda=50$  the distribution $S_n$ eventually develops the exponential decay for $n\gg 50$, a 'power-law-like' region suggests that in the special conditions of long-range spatial correlations the alpha stable distributions might be a more relevant description of the chain statistics than the Gaussian distribution. For comparison we also include in Fig. \ref{fig:linearized} the data from \cite{bib:majka1}, which preserve the exponential form in the entire range of parameters. 

\section{Alpha-stable chain model}\label{sec:model}
In the Gaussian chain model, the geometry of a chain is described as a random walk trajectory, in which the distribution of the distances between nearest neighbors is Gaussian, namely \cite{bib:teraoka}:
\begin{equation}
G(|\textbf{r}_{i+1}-\textbf{r}_i|)=\left(\frac{2\pi b^2}{D} \right)^{-D/2}\exp\left(-\frac{D(\textbf{r}_{i+1}-\textbf{r}_i)^2}{2b^2} \right)
\end{equation}
Here, $D$ is the dimension of the system, $b$ is usually interpreted as the length of a segment and $\textbf{r}_i$ is the vector position of $i$-th node. The characteristic function of $G$ reads:
\begin{equation}
\phi_{G}(k)=\exp\left(-\frac{2b^2}{D}k^2 \right) \label{eq:gauss}
\end{equation}

Let us now consider the generalization of $G(|\textbf{r}_{i+1}-\textbf{r}_i|)$ to the alpha-stable distribution $P(|\textbf{r}_{i+1}-\textbf{r}_i|)$. The multivariate alpha-stable distributions are defined in terms of their characteristic functions, which can be written in the following parametrization \cite{bib:samorodnitsky}:
\begin{widetext}
\begin{equation}
\phi(\textbf{k})=
\begin{aligned}
\exp\left(-\int_{S_D}|\textbf{k}\cdot\textbf{s}|^{\alpha}\left(1-i \textrm{sgn}(\textbf{k}\cdot\textbf{s})\tan\frac{\pi\alpha}{2}\right)\Gamma(d\textbf{s})+i\textbf{k}\cdot \boldsymbol \mu\right)&&\textrm{for}&& \alpha\neq1\\
\exp\left(-\int_{S_D}|\textbf{k}\cdot\textbf{s}|\left(1+i  \textrm{sgn}(\textbf{k}\cdot\textbf{s})\frac{2\ln\textbf{k}\cdot\textbf{s}}{\pi}\right)\Gamma(d\textbf{s})+i\textbf{k}\cdot\boldsymbol \mu \right)&&\textrm{for}&& \alpha=1
\end{aligned}
\end{equation}
\end{widetext}
In the above definition $\Gamma(d\textbf{s})$ stands for the spectral measure defined on the $D$-dimensional unit sphere $S_D$, $\textbf{k}\cdot\textbf{s}$ denotes the scalar product and $\boldsymbol \mu$ is the vector mean value. Since we are interested in the spherically symmetric distributions, we choose the uniform spectral measure $\Gamma(d\textbf{s})=const$ \cite{bib:dybiec}. Under such choice, and assuming $\boldsymbol \mu=0$, the general parametrization can be simplified to the following form:
\begin{equation}
\phi(k)=\exp \left(-c k^{\alpha} \right) \label{eq:char_fun}
\end{equation}
where $k=|\textbf{k}|$ and $c$ is a constant. Eq. \eqref{eq:gauss} is a special case of $\phi(k)$ and our choice of $c$ should agree with $\phi_G(k)$ for $\alpha=2$. Therefore, we postulate that $c$ reads:
\begin{equation}
c=\frac{2b^{\alpha}}{D}
\end{equation}
and, finally, the nearest neighbor spatial distribution in the alpha-stable chain model reads:
\begin{equation}
P(|\textbf{r}_{i+1}-\textbf{r}_i|)=\frac{1}{(2\pi)^D}\int d\textbf{k} \exp \left(i\textbf{k}\cdot (\textbf{r}_{i+1}-\textbf{r}_i)-\frac{2}{D}b^{\alpha}k^{\alpha}\right) \label{eq:P_def}
\end{equation}

Having established the nearest neighbors spatial distribution $P(|\textbf{r}_{i+1}-\textbf{r}_i|)$, we can calculate such distribution for any pair of segments, namely:
\begin{equation}
\begin{split}
&P(|\textbf{r}_{i}-\textbf{r}_j|)=\\
&=\int d\textbf{r}_{i+1}\dots \int d\textbf{r}_{j-1}\prod_{n=i+1}^{j} P(|\textbf{r}_{n}-\textbf{r}_{n-1}|)=\\
&=\frac{1}{(2\pi)^D}\int d\textbf{k} \exp\left(i\textbf{k}\cdot(\textbf{r}_i-\textbf{r}_j)-\frac{2}{D}|i-j|b^{\alpha}k^{\alpha} \right)
\end{split}
\end{equation}
where we made use of alpha-stability. For $\alpha=2$ this formula comes down to the well known result for Gaussian chain \cite{bib:teraoka}:
\begin{equation}
G(|\textbf{r}_{i}-\textbf{r}_j|)=\left( \frac{2\pi|i-j|b^2}{D}\right)^{-D/2} \exp \left( -\frac{D(\textbf{r}_i-\textbf{r}_j)^2}{2|i-j|b^2}\right)
\end{equation}
It is also possible to calculate $\alpha=1$ case explicitly:
\begin{equation}
P_{\alpha=1, D=3}(|\textbf{r}_{i}-\textbf{r}_j|)=\frac{16\pi |i-j|}{3}\frac{1}{\left(\frac{4|i-j|^2}{9}+r^2\right)^2}
\end{equation}
In particular, taking as $i$ and $j$ the first and the last segment respectively, we obtain the end-to-end distance distribution.

A classical problem in polymer physics is to predict the scaling behavior of radius of gyration $R_g$ with growing $N$. In order to do so, we will calculate $R_g$ using a method mentioned in \cite{bib:moon}, namely  $R_g=<r^{\alpha}>^{1/\alpha}$, where $<.>$ denotes the average: 
\begin{equation}
\begin{split}
&R_g=\\
&=\left(\frac{1}{(2\pi)^D}\int d\textbf{r}r^{\alpha}\int d\textbf{k}\exp\left(i\textbf{k}\cdot \textbf{r}-\frac{2N}{D} b^{\alpha}k^{\alpha} \right) \right)^{1/\alpha} \\
&=b\left(\frac{2N}{D}\right)^{1/\alpha}\left(\frac{1}{(2\pi)^D}\int d\textbf{r}'r'^{\alpha}\int d\textbf{k}'\exp\left(i\textbf{k}'\cdot \textbf{r}'-k'^{\alpha} \right) \right)^{1/\alpha}
\end{split} \label{eq:Rg}
\end{equation}
This result is obtained via the change of variables $\textbf{k}'=b(2N/D)^{1/\alpha}\textbf{k}$ and $\textbf{r}'=b^{-1}(2N/D)^{-1/\alpha}\textbf{r}$, which completely factors the dependence on $b$ and $N$ out of the integral. Therefore, the scaling reads $R_g\propto b N^{1/\alpha}$.

\section{Distribution of segments around the center of mass}\label{sec:mass_center}
The central problem of deriving the coarse-grained interaction between two chains is to find the distribution of segments around the mass center of a single chain. Let us consider an $N$-segments long chain, described by the nearest neighbor probability given by \eqref{eq:P_def}. The position of the mass center reads:
\begin{equation}
\textbf{R}=\frac{1}{N}\sum_{i=1}^{N}\textbf{r}_i
\end{equation}
The probability that each segment occupies its position $\textbf{r}_i$ under condition that the mass center is positioned at $\textbf{R}$ reads:
\begin{equation}
P(\textbf{r}_1,\dots,\textbf{r}_N|\textbf{R})=\prod_{i=1}^{N-1}P(|\textbf{r}_{i+1}-\textbf{r}_i|)\delta \left(\textbf{R}-\frac{1}{N}\sum_{i=1}^{N}\textbf{r}_i\right)
\end{equation}
From this expression we can calculate the probability of finding $j$-th segment at some position relative to the mass center:
\begin{equation}
\begin{split}
P&(|\textbf{r}_j-\textbf{R}|)=\int d\textbf{r}_1\dots\int d\textbf{r}_{j-1}\int d\textbf{r}_{j+1}\dots \int d\textbf{r}_{N}\times \\ &\times\prod_{i=1}^{N-1}P(|\textbf{r}_{i+1}-\textbf{r}_i|)\delta \left(\textbf{R}-\frac{1}{N}\sum_{i=1}^{N}\textbf{r}_i\right)\label{eq:P1}
\end{split}
\end{equation}
The integrals in the above expression can be done particularly easily, if we switch to relative variables:
\begin{equation}
\begin{aligned}
\Delta \textbf{r}_{i-j}=\textbf{r}_{i}-\textbf{r}_{i-1}&&\textrm{for}&&i> j\\
\Delta \textbf{r}_{i-j}=\textbf{r}_{i-1}-\textbf{r}_{i}&&\textrm{for}&&i<j \label{eq:new_var}
\end{aligned}
\end{equation}
which allows us to express $\textbf{r}_i$ as:
\begin{equation}
\begin{aligned}
\textbf{r}_i=\textbf{r}_j+\sum_{n=1}^{N-i}\Delta \textbf{r}_{+n}&&\textrm{for}&&N\ge i>j \\
\textbf{r}_i=\textbf{r}_j+\sum_{n=1}^{j-i}\Delta \textbf{r}_{-n}&&\textrm{for}&&1\le i<j
\end{aligned}
\end{equation}
In these coordinates the position of mass center reads:
\begin{equation}
\frac{1}{N}\sum_{i=1}^N\textbf{r}_i=\textbf{r}_j+\sum_{n=1}^{N-j}\frac{N-j-n+1}{N}\Delta \textbf{r}_{+n}+\sum_{n=1}^{j-1}\frac{j-n}{N}\Delta\textbf{r}_{-n}
\end{equation}
The change of variables \eqref{eq:new_var} is linear and its Jacobian is equal 1, so applying the new coordinates to \eqref{eq:P1}, we obtain:
\begin{equation}
\begin{split}
&P(|\textbf{r}_j-\textbf{R}|)=\prod_{\substack{n=-j\\n\neq 0}}^{N-j}\int d\Delta\textbf{r}_n P(\Delta r_n)\times\\
&\times\delta\left(\textbf{R}- \textbf{r}_j+\sum_{n=1}^{N-j}\frac{N-j-n+1}{N}\Delta \textbf{r}_{+n}+\sum_{n=1}^{j-1}\frac{j-n}{N}\Delta\textbf{r}_{-n}\right) \label{eq:P2}
\end{split}
\end{equation}
The following step is to express $P(\Delta r_n)$ in \eqref{eq:P2} in terms of its characteristic function \eqref{eq:P_def}:
\begin{equation}
\begin{split}
&P(|\textbf{r}_j-\textbf{R}|)=\frac{1}{(2\pi)^{ND}}\prod_{\substack{n=-j\\n\neq 0}}^{N-j}\int d\Delta\textbf{r}_n\int d \textbf{k}_n e^{i\textbf{k}_n\cdot \Delta \textbf{r}_n}\phi(k_n)\times \\
 &\times\int d\textbf{k}_0 \exp\Biggl( i\textbf{k}_0\cdot(\textbf{R}-\textbf{r}_j)+\Biggr.\\
&\left.+i \sum_{n=1}^{N-j}\frac{N-j-n+1}{N}\textbf{k}_0\cdot \Delta \textbf{r}_{+n}+i\sum_{n=1}^{j-1}\frac{j-n}{N}\textbf{k}_0\cdot\Delta\textbf{ r}_{-n}\right)
\end{split}
\end{equation}
Further, we integrate out every component of $\Delta\textbf{r}_{\pm n}$, which introduces multiple Dirac-delta functions:
\begin{equation}
\begin{split}
P&(|\textbf{r}_j-\textbf{R}|)=\frac{1}{(2\pi)^{D}}\times\\
=&\int d\textbf{k}_0\left(\prod_{n=1}^{N-j}\int d\textbf{k}_{+n}\phi(k_{+n})\delta\left( \textbf{k}_{+n}-\frac{N-j-n+1}{N}\textbf{k}_0\right) \right)\times\\
&\times\left(\prod_{n=1}^{j-1}\int d\textbf{k}_{-n}\phi(k_{-n})\delta\left(\textbf{k}_{-n}-\frac{j-n}{N}\textbf{k}_0\right) \right) e^{i\textbf{k}_0\cdot(\textbf{R}-\textbf{r}_j)}=\\
=&\frac{1}{(2\pi)^{D}}\int d\textbf{k}_0 \left( \prod_{n=1}^{N-j}\phi \left(\frac{N-j-n+1}{N}k_0\right)\right)\times\\
&\times\left( \prod_{n=1}^{j-1}\phi\left(\frac{j-n}{N}k_0\right)\right)e^{i\textbf{k}_0\cdot(\textbf{R}-\textbf{r}_j)}
\end{split}
\end{equation}
At this point we apply the explicit form of $\phi(k)$, so the final expression for $P(|\textbf{r}_j-\textbf{R}|)$ reads:
\begin{equation}
\begin{split}
&P(|\textbf{r}_j-\textbf{R}|)=\frac{1}{(2\pi)^{D}}\int d\textbf{k}_0 \exp\Biggl[ i\textbf{k}_0\cdot(\textbf{R}-\textbf{r}_j)-\Biggr. \\
&\left.-\frac{2b^{\alpha}}{D} \left(\sum_{n=1}^{N-j}\left(\frac{N-j-n+1}{N}\right)^\alpha+ \sum_{n=1}^{j-1}\left(\frac{j-n}{N}\right)^\alpha\right)k_0^\alpha\right] \label{eq:P3}
\end{split}
\end{equation}
Expression \eqref{eq:P3} gives the probability of finding $j$-th segment in the vicinity of mass center, so the probability of finding any segment reads:
\begin{equation}
P_{CM}(|\textbf{r}-\textbf{R}|)=\frac{1}{N}\sum_{j=1}^{N} P(|\textbf{r}_j-\textbf{R}|)
\end{equation}
where the factor $1/N$ provides normalization. Let us assume now that $N$ is a large number, so both $n/N=q$ and $j/N=q'$ can be treated as continuous variables, hence we can simplify:
\begin{gather}
\begin{split}
\sum_{n=1}^{N-j}\left(\frac{N-j-n+1}{N}\right)^\alpha&\to N\int_0^{1-q'}dq(1-q'-q)^\alpha \\
&=\frac{N}{\alpha+1}(1-q')^{\alpha+1} 
\end{split}
\\
 \sum_{n=1}^{j-1}\left(\frac{j-n}{N}\right)^\alpha\to N\int_0^{q'}dq (q'-q)^{\alpha}=\frac{N}{\alpha+1}q'^{\alpha+1}
\end{gather}
The final expression for the distribution of any segment around the mass center reads:
\begin{equation}
\begin{split}
&P_{CM}(|\textbf{r}-\textbf{R}|)=\frac{1}{(2\pi)^{D}}\int_0^1dq' \int d\textbf{k}_0 \times\\
&\times\exp\left[i\textbf{k}_0\cdot(\textbf{r}-\textbf{R})- \frac{2Nb^{\alpha}k_0^{\alpha}}{D(\alpha+1)}((1-q')^{\alpha+1}+q'^{\alpha+1}) \right]\label{eq:P4}
\end{split}
\end{equation}
While this expression is exact, the integral with respect to $q'$ makes it unwieldy. We can simplify it by resorting to the integral mean value theorem. Namely, there exists such $q'_0\in [0,1]$, that:
\begin{equation}
\begin{split}
&P_{CM}(|\textbf{r}-\textbf{R}|)=\frac{1}{(2\pi)^{D}}\int d\textbf{k}_0 \times \\
&\exp\left[i\textbf{k}_0\cdot(\textbf{r}-\textbf{R})- \frac{2Nb^{\alpha}k_0^{\alpha}}{D(\alpha+1)}((1-q_0')^{\alpha+1}+q_0'^{\alpha+1}) \right] \label{eq:P5}
\end{split}
\end{equation}
On the other hand, since the integrand of \eqref{eq:P4} is a peak function of $q'$ with maximum at $q'=1/2$, this value contributes the most to the integral. For this reason, we will approximate \eqref{eq:P5} by imposing $q'_0=1/2$. Once again, this results in the characteristic function of $P_{CM}(|\textbf{r}-\textbf{R}|)$ in a form $\exp(-c(\alpha)k^{\alpha})$, where:
\begin{equation}
c(\alpha)=\frac{2N}{D(\alpha+1)}\left(\frac{b}{2} \right)^{\alpha}
\end{equation}

\section{Coarse grained interaction between two non-Gaussian polymers}\label{sec:interaction}
Having established $P_{CM}(|\textbf{r}-\textbf{R}|)$ for a single chain, we can calculate the coarse-grained interaction between two chains in terms of the distance between their mass centers. From now on, lower index numerates the type of particle, so $P_{CM,i}(|\textbf{r}-\textbf{R}_i|)$ describes an $i$-th type of chain characterized by $N_i$ segments, constant $b_i$ and exponent $\alpha_i$. We can follow the reasoning of Flory \cite{bib:flory} and assume that the systems suffers an energetic penalty $\epsilon_{ij}$ if a segment belonging to the first chain invades a small volume in the vicinity of a segment belonging to the other chain. For a single site $\textbf{r}$, the probability of such event is proportional to $P_{CM,i}(|\textbf{r}-\textbf{R}_i|)P_{CM,j}(|\textbf{r}-\textbf{R}_j|)d\textbf{r}$. Therefore, the entire interaction reads:
\begin{equation}
\begin{split}
&V_{ij}(|\textbf{R}_i-\textbf{R}_j|)=\\
&=\epsilon_{ij}\tilde c_{ij}\int d\textbf{r}P_{CM,i}(|\textbf{r}-\textbf{R}_i|)P_{CM,j}(|\textbf{r}-\textbf{R}_j|)=\\
&=\frac{\epsilon_{ij}\tilde c_{ij}}{(2\pi)^D}\int d\textbf{k} \exp \left(i\textbf{k}\cdot(\textbf{R}_i-\textbf{R}_j)-c_i(\alpha_i)k^{\alpha_i}-c_j(\alpha_j)k^{\alpha_j} \right)\label{eq:pot1}
\end{split}
\end{equation}
Assuming that $\epsilon_{ij}$ has a dimension of energy, it is necessary to introduce an additional constant $\tilde c_{ij}$, which has a dimension of volume. We can deduce this constant from the case of $\alpha_i=\alpha_j=2$, for which we obtain the following 'universal' Gaussian potential \cite{bib:likos} and its Fourier transform: 
\begin{align}
V(r)=\epsilon\exp\left(-\frac{r^2}{4 c} \right)&&\mathcal{V}(k)=\epsilon  \left(4\pi c\right)^{D/2}e^{- c k^2} \label{eq:gauss_pot}
\end{align}
When $\epsilon$ is independent from $N$, this potential is perceived as an accurate and reliable model for interaction of identical chains \cite{bib:likos, bib:bolhuis}. Comparing \eqref{eq:gauss_pot} to \eqref{eq:pot1}, one can see that $c=c_i(2)+c_j(2)$ and hence $\tilde c_{ij}=(4\pi c)^{D/2}$. This can be generalized for $\alpha_i=\alpha_j=\alpha$ by:
\begin{equation}
\tilde c_{ij}=\left(4\pi(c_i(\alpha)+c_j(\alpha)) \right)^{D/\alpha}\label{eq:c_ij}
\end{equation}
For the case of $\alpha_i\neq\alpha_j$ the constant $\tilde c_{ij}$ cannot be uniquely deduced from the dimensional analysis, thus we will restrict our further considerations to the potentials with common $\alpha$. 

\section{Effective interactions and mixture stability}\label{sec:bin_mix}
Once $V_{ij}$ has been found, we can analyze the interactions in binary mixtures. The system is described by three microscopic potentials in the form \eqref{eq:pot1}, where $V_{11}$ and $V_{22}$ are the internal interactions of each species and $V_{12}$ is the cross-species interaction. When the behavior of one species in a mixture is considered, the presence of the other species modifies the microscopic interaction \cite{bib:lekkerkerker, bib:likos}. The additional potential, known as the effective interaction, is of entropic origin \cite{bib:lekkerkerker, bib:likos} and it is a key-factor in controlling mixture stability and demixing. The prediction of effective interactions from arbitrary microscopic potentials is usually a challenging numerical task, but in our previous work  \cite{bib:eff_int} we have proposed a simple analytical method, suitable for soft interactions. According to \cite{bib:eff_int}, the effective interaction can be estimated by:
\begin{equation}
U_{eff}(\Delta R)=-\frac{1}{(2\pi)^D}\int_{\tilde \Omega} d\textbf{k} e^{i\textbf{k}\cdot\Delta\textbf{R}}\frac{|\mathcal{V}_{12}(k)|^2}{\mathcal{V}_{22}(k)}
\end{equation}
where $\mathcal{V}_{ij}(k)=\int d\textbf{r}\exp(i\textbf{k}\cdot\textbf{r})V_{ij}(r)$ is a Fourier transform of relevant $V_{ij}(r)$ and $\tilde \Omega$ is the volume in the reciprocal space. Substituting \eqref{eq:pot1} with relevant constants given by \eqref{eq:c_ij} into the expression for effective interactions, one obtains:
\begin{equation}
\begin{split}
U_{eff}(\Delta R)=&-\frac{1}{(2\pi)^D}\frac{\epsilon_{12}^2}{\epsilon_{22}}\frac{(4\pi)^{D/\alpha}(c_1(\alpha)+c_2(\alpha))^{2D/\alpha}}{( 2c_2(\alpha))^{D/\alpha}}\times\\
&\times\int_{\tilde \Omega} d\textbf{k} e^{i\textbf{k}\cdot\Delta\textbf{R}-2c_1(\alpha)k^{\alpha}}
\end{split}
\end{equation}
The total interaction for the first species in the mixture reads:
\begin{equation}
U_{tot}(\Delta R)=V_{11}(\Delta R)+U_{eff}(\Delta R) \label{eq:U_tot}
\end{equation}
or explicitly:
\begin{equation}
\begin{split}
&U_{tot}(r)=\\
&\left(\epsilon_{11}\left( 4\pi c_1(\alpha)\right)^{D/\alpha}-\frac{\epsilon_{12}^2}{\epsilon_{22}}\frac{(4\pi)^{D/\alpha}(c_1(\alpha)+c_2(\alpha))^{2D/\alpha}}{(2c_2(\alpha))^{D/\alpha}}\right)\times\\
&\times\frac{1}{(2\pi)^D} \int_{\tilde \Omega} d\textbf{k} e^{i\textbf{k}\cdot\Delta\textbf{R}-2c_1(\alpha)k^{\alpha}}
\end{split}
\end{equation}

One can see that $U_{tot}(\Delta R)$ and $V_{11}(\Delta R)$ have the same shape, up to the scaling constant $S$:
\begin{equation}
S=\epsilon_{11}\left( 4\pi c_1(\alpha)\right)^{D/\alpha}-\frac{\epsilon_{12}^2}{\epsilon_{22}}\frac{(4\pi)^{D/\alpha}(c_1(\alpha)+c_2(\alpha))^{2D/\alpha}}{(2c_2(\alpha))^{D/\alpha}}
\end{equation}
S has a complicated form and it can take both the negative and positive values, depending on the parameters. The change in the sign of the total interaction indicates a remarkable change in the behavior of the system. Namely, for $S>0$ the interaction between the particles of the first species are purely repulsive, which means that these particles will disperse in the volume. Conversely, for $S<0$, the first species of particles interacts via attractive potential, which results in the clustering of these particles and demixing in the system. Equating $S$ to 0, the condition for demixing reads:
\begin{equation}
\frac{\epsilon_{11}\epsilon_{22}}{\epsilon_{21}^2}<\left(\frac{(c_1(\alpha)+c_2(\alpha))^{2}}{4 c_1(\alpha) \label{eq:cond} c_2(\alpha)}\right)^{D/\alpha}
\end{equation}

Let us now analyze the condition \eqref{eq:cond} and introduce a common energy scale:
\begin{equation}
\tilde \epsilon=\frac{\epsilon_{12}}{\sqrt{\epsilon_{11}\epsilon_{22}}}
\end{equation}
 and: 
\begin{equation}
g=\left( \frac{c_1(\alpha)}{c_2(\alpha)} \right)^{1/\alpha}=\frac{b_1}{b_2}\left(\frac{N_1}{N_2}\right)^{1/\alpha}
\end{equation}
for which condition \eqref{eq:cond} can be reduced to:
\begin{equation}
\tilde \epsilon>\left(\frac{4g^{\alpha}}{(1+g^{\alpha})^2} \right)^{D/(2\alpha)} \label{eq:cond_fin}
\end{equation}
Recalling the equation \eqref{eq:Rg} for the radius of gyration $R_g$, one can see that for chains characterized by the distributions sharing the same $\alpha$, the parameter $g$ becomes the ratio of $R_g$:
\begin{equation}
g=\frac{R_{g,1}}{R_{g,2}}
\end{equation}
For $\alpha=2$ and $D=3$ the condition \eqref{eq:cond} becomes exactly the spinodal decomposition condition for Gaussian particles, as given in \cite{bib:sep1} and \cite{bib:sep2}, namely:
\begin{equation}
\tilde \epsilon>\left(\frac{2g}{1+g^2}\right)^{3/2}
\end{equation}
Therefore, the condition \eqref{eq:cond_fin} is a direct generalization of the spinodal decomposition to the systems of particles described with alpha-stable distributions. 

\begin{figure}
\includegraphics[width=0.95\columnwidth]{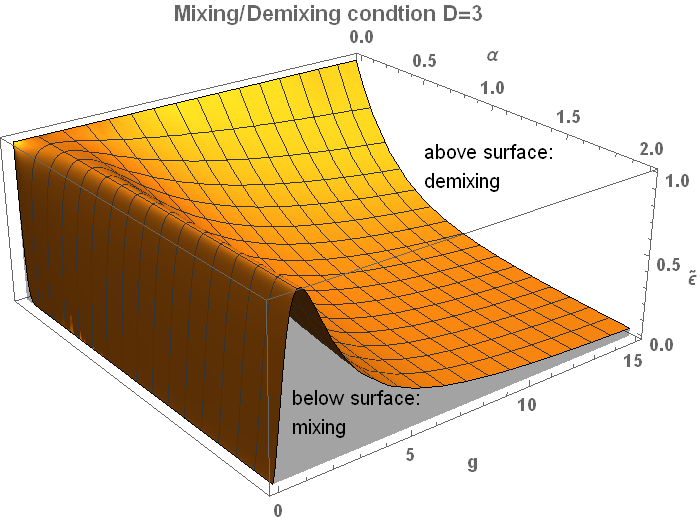}
\caption{The spinodal decomposition condition \eqref{eq:cond_fin} for the binary systems of particles described with the alpha-stable statistics as a function of gyration radii ratio $g=R_{g,1}/R_{g,2}$ and the distribution exponent $\alpha$ for $D=3$ dimensional system. $\tilde \epsilon=\epsilon_{12}/\sqrt{\epsilon_{11}\epsilon_{22}}$ is the common energy scale. In the region above the plotted surface the binary system undergoes demixing due to the prevalence of the attractive effective interactions, in the region below the surface the effective interaction is repulsive and the system is homogeneous.} \label{fig:demixing}
\end{figure}

The condition \eqref{eq:cond_fin} is plotted in Fig. \ref{fig:demixing}. For every pair of $g$ and $\alpha$ its value fits between 0 and 1. In the entire range of $\alpha$, the region of mixing (below the surface) preserves the features of the Gaussian case, namely it falls rapidly to 0 for $g\ll1$, reaches the single maximum at $g=1$ and asymptotically decreases to 0 for $g\gg1$. However, as $\alpha$ decreases to 0 the mixing region for $g\gg1$ becomes wider, asymptotically reaching the region defined by $\tilde\epsilon>1$. This means that the gyration radius ratio $g$ becomes less and less relevant for the mixing of chains characterized by very wide distributions. The changes in the mixing region shape are much more pronounced for $\alpha<1$.

\section{Discussion}\label{sec:discussion}
It is known that the spinodal decomposition condition for Gaussian particles leads to the predictions on mixture separation which are qualitatively and quantitatively comparable to the more advanced methods \cite{bib:sep2}. Thus, a similar efficiency can be expected from \eqref{eq:cond_fin}, at least for $\alpha$ mildly deviating from 2. However, some possible issues should be commented.

The fact that the total interaction \eqref{eq:U_tot} can become entirely negative is unrealistic. This evident problem is mitigated by the fact, that the energy density of a pair interaction behaves as $U_{tot}(r) r^{D-1}dr$. Therefore, $r^{D-1}$ factor suppresses the lack of repulsive core at short distances and amplifies the influence of the attractive tail. The unrealistic shape of $U_{tot}$ is also a consequence of the way the potential $V_{ij}(r)$ given by \eqref{eq:pot1} is constructed. This potential is mean-field in its nature and its width is governed by the constant $c_i(\alpha)+c_j(\alpha)$. In the context of Gaussian particles, while the dominant shape of the interaction between two separate chains is agreed to be Gaussian \cite{bib:bolhuis, bib:mc1}, there is an open problem of whether there are additional components \cite{ bib:mc1} or how the width of such Gaussian is related to the gyration radii of the component chains \cite{bib:sep1}. A similar problem is relevant in our case and the choice of the width constant different from $c_i(\alpha)+c_j(\alpha)$ might result in a more realistic shape of $U_{tot}$

\section{Phase separation in the adsorption of Gaussian particles to the surface}\label{sec:gauss_sep}
The result \eqref{eq:cond_fin} is particularly interesting in the context of the already mentioned work of Bouchaud and Daoud \cite{bib:bouchaud}, where the gyration radius parallel to the surface is calculated for an adsorbed polymer. As mentioned in Section \ref{sec:interpretation}, the characteristic exponent for the distribution on the surface reads $\alpha_{ss}=2$ in the strong adsorption limit and $\alpha_{ws}=1$ in the weak adsorption limit \cite{bib:bouchaud}. Considering the adsorption from the binary mixture, \eqref{eq:cond_fin} provides the condition for homogeneous versus inhomogeneous adsorption, i. e. the ratio of gyration radii parallel to the surface ($R_{g,||}$) decides whether both species cover the surface in a homogeneous manner or they separate into the islands consisting of the particles of the same type. On the other hand \eqref{eq:cond_fin} allows us to compare for which parameters the separation on the surface and in the bulk co-appear. 

Let us consider a simple binary system in which the behavior of chains in the bulk ($D_b=3$, $\alpha_b=2$) is Gaussian, but on the surface it is characterized by $D_s=2$ and $\alpha_{ss}$ or $\alpha_{ws}$. The types of particles differ by the number of monomers $N_i$ and their persistence length $b_i$. The condition \eqref{eq:cond} reads:
\begin{equation}
\tilde \epsilon > \left( \frac{4\left( \frac{b_1}{b_2}\right)^{\alpha_{x}}\frac{N_1}{N_2}}{\left(1+\left( \frac{b_1}{b_2}\right)^{\alpha_{x}}\frac{N_1}{N_2}\right)^2}\right)^{\frac{D_y}{2\alpha_x}}=f_{x,y} \label{eq:cond_sep}
\end{equation}
In the strong adsorption limit it is always true that $f_{ss, s}\ge f_{b,b}$ for any $b_1/b_2$ and $N_1/N_2$. Therefore, assuming that in this system $\tilde \epsilon$ is the same on the surface and in the bulk, three scenarios of mixing/demixing are allowed. First, for (i) $f_{ss,s}\ge f_{b,b}\ge\tilde \epsilon$ the solution in the bulk is homogeneous and so is the coverage of the surface. Conversely, for (ii) $\tilde\epsilon\ge f_{ss,s}\ge f_{b,b}$ the separation is simultaneous on the surface and in the bulk. Finally, for (iii) $f_{ss,s}\ge\tilde \epsilon\ge f_{b,b}$  demixing in the bulk occurs, but the surface coverage is still homogeneous. 

\begin{figure}
\includegraphics[width=0.95\columnwidth]{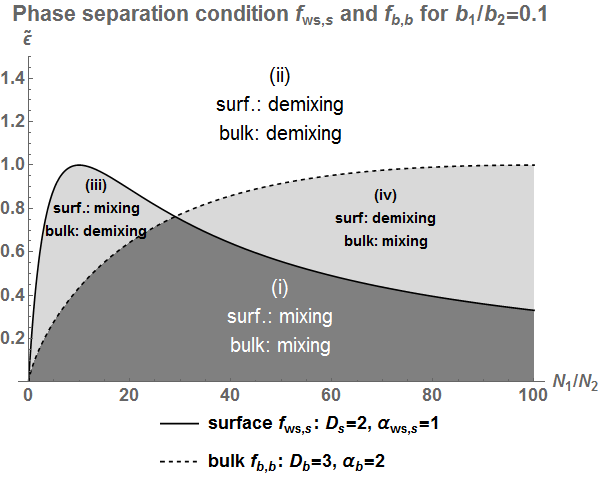}
\caption{The exemplary comparison of the phase separation conditions on the surface in the weak adsorption limit ($f_{ws,s}$, solid line) and in the bulk ($f_{b,b}$, dashed line) for the binary system of Gaussian polymers characterized by the persistence length ratio $b_1/b_2=0.1$. $f_{ws,s}$ and $f_{b,b}$ are defined by \eqref{eq:cond_sep}. $f_{ws,s}$ and $f_{b,b}$ divide the plot into four regions: (i) simultaneous mixing on the surface and in the bulk, (ii) simultaneous demixing on the surface and in the bulk, (iii) mixing on the surface, demixing in the bulk, (iv) demixing on the surface, mixing in the bulk. \label{fig:phase1}}
\end{figure}

\begin{figure}
\includegraphics[width=0.98\columnwidth]{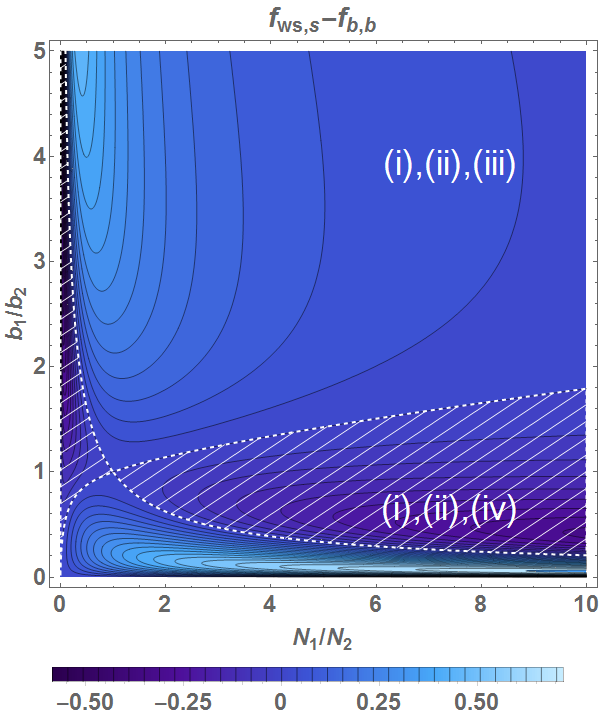}
\caption{The difference $f_{ws,s}-f_{b,b}$ indicates which surface/bulk demixing scenarios are allowed (see Section \ref{sec:gauss_sep} and Fig. \ref{fig:phase1} for explanation). White meshed region: $f_{ws,s}-f_{b,s}<0$ indicates scenario (iii) allowed (demixing on the surface, mixing in the bulk), complement region: $f_{ws,s-}f_{b,b}>0$ scenario (iv) allowed (mixing on the surface, demixing in the bulk). \label{fig:phase2}}
\end{figure}

In the weak adsorption limit ($\alpha_{ws}=1$), the situation is more complicated because both $f_{ws,s}\ge f_{b,b}$ and $f_{ws,s}\le f_{b,b}$ are possible, depending on $b_1/b_2$, $N_1/N_2$. Replacing $f_{ss,s}$ by $f_{ws,s}$ in the inequalities from the previous paragraph one obtains the conditions for separation scenarios (i)-(iii) in the weak adsorption limit. However, there exists the additional region in which it is possible that (iv) $f_{b,b}\ge\tilde\epsilon \ge f_{ws,s}$. In this region the separation on the surface  occurs, while the solution in the bulk is still homogeneous. In Fig. \ref{fig:phase1} the exemplary phase separation diagram for the weak adsorption limit and $b_1/b_2=0.1$ is presented, which contains all of the phase separation scenarios (i)-(iv). Fig. \ref{fig:phase2} shows the difference $f_{ws,s}-f_{b,b}$, which indicates where scenarios (iii) and (iv) are allowed. In particular, for $b_1/b_2\to0$ and $b_1/b_2\gg1$ the scenario (iv) becomes almost inaccessible, while it is allowed in the vicinity of $b_1/b_2\simeq1$.

While these considerations show that the behavior of mixture can be designed by the choice of $b_1/b_2$ (which is dependent on the chemical composition) and $N_1/N_2$, some additional effects might affect the result. One issue is that there exists the additional entropic attraction between surface and the bigger particles \cite{bib:lekkerkerker, bib:yodh} which is not included here. Another problem is that our considerations are relevant for the thermodynamic equilibrium, which might be difficult to achieve in the adsorption process. 


\section{Summary}\label{sec:summary}
In summary, in this paper we have presented the generalization of the results known from the Gaussian chain theory to the particles described with the alpha-stable distributions. As expected, it is possible to obtain a similar hierarchy of analytical results ranging from end-to-end distribution up to the effective interactions in binary mixtures. Typically for alpha-stable distributions, we obtained the closed-form formulas in the Fourier space. Our theory also allows us to generalize the spinodal decomposition condition from Gaussian particles to the alpha-stable particles. This can be readily applied to the problem of mixing/demixing of adsorbed polymers, as we also show. Our results might be further utilized in the context of Levy flights applications reviewed in Section \ref{sec:interpretation}.

\end{document}